\documentclass[lettersize,journal]{IEEEtran}
\usepackage{amsmath,amsfonts}
\usepackage{algorithm,algpseudocode}
\usepackage{array}
\usepackage[labelsep=period]{caption}
\usepackage[belowskip=-5pt,aboveskip=10pt,font=footnotesize, justification=raggedright, singlelinecheck=false]{caption}
\usepackage{textcomp}
\usepackage{stfloats}
\usepackage{url}
\usepackage{verbatim}
\usepackage{graphicx}
\usepackage{amssymb}
\usepackage{kotex}
\usepackage{cite}
\usepackage{setspace}

\begin{document}

\title{AIRS-assisted Vehicular Networks with Rate-Splitting SWIPT Receivers: Joint Trajectory and Communication Design}

\author{Gyoungyoon Nam, Seokhyun Lee and Seongah Jeong,~\IEEEmembership{Member,~IEEE}
\thanks{This research was supported by Kyungpook National University Research Fund, 2022. \\
\indent Gyoungyoon Nam, Seokhyun Lee and Seongah Jeong are with the School of Electronics Engineering, Kyungpook National University, Daegu 14566, South Korea (e-mail: ngy3827yes@knu.ac.kr, kei04060@knu.ac.kr, seongah@knu.ac.kr) }
}

\markboth{Journal of \LaTeX\ Class Files,~Vol.~14, No.~8, August~2021}%
{Shell \MakeLowercase{\textit{et al.}}: A Sample Article Using IEEEtran.cls for IEEE Journals}

\maketitle

\begin{abstract}
In this correspondence, we propose to use an intelligent reflective surface (IRS) installed on unmanned aerial vehicle (UAV), referred to as aerial IRS (AIRS), for vehicular networks, where simultaneous wireless information and power transfer (SWIPT) receivers to concurrently allow information decoding (ID) and energy harvesting (EH) are equipped at the battery-limited vehicles. For efficiently supporting the multiple moving vehicles, we adopt rate-splitting multiple access (RSMA) technique. With the aim of maximizing the sum rate of vehicles, we jointly optimize trajectory and phase shift design of AIRS, transmit power and rate allocation for RSMA along with power splitting ratio for SWIPT implementation. Via simulations, the superior performances of the proposed algorithm are validated compared to the conventional partial optimizations.
\end{abstract}


\section{Introduction}
\IEEEPARstart{W}{ith} the explosive growth in wireless mobile communications, vehicle-to-everything (V2X) services have been actively explored to exchange messages with other vehicles, pedestrians and infrastructures for road safety, traffic control and various traffic notifications. In order to lengthen the lifetime of vehicular networks, simultaneous wireless information and power transfer (SWIPT) has received much attention for the battery-limited vehicles \cite{VANETandSWIPT1, VANETandSWIPT2}. For the SWIPT implementation, either signal power or time can be leveraged by splitting for information decoding (ID) and energy harvesting (EH), the former and the latter of which are called as power splitting (PS) and time switching (TS) method, respectively. In \cite{VANETandSWIPT1, VANETandSWIPT2}, the SWIPT technology has been investigated to cope with the dynamic vehicular channels, e.g., by considering the channel imperfectness or selective characteristics.

\indent For stability and reliability in vehicular communication environments with the high-speed mobility, terrestrial or aerial intelligent reflecting surface (IRS) with a number of cost-effective passive reflecting elements has been explored \cite{AIRS, mainRef}. Specifically, thanks to the high possibility of enabling the indirect transmission for dark zones of access point (AP), the IRS mounted on unmanned aerial vehicles (UAV), referred to as aerial IRS (AIRS), has been proposed with the flexible mobility of in three-dimension (3D) space. Moreover, the authors in \cite{VANETandAIRS} propose the design to maximize the average bit rate of vehicular networks by using the AIRS to alleviate the battery capacity of vehicles.

\indent Motivated by these previous works \cite{VANETandSWIPT1, VANETandSWIPT2, VANETandAIRS}, we propose the AIRS-assisted vehicular networks, where the vehicles are equipped with the rate-splitting SWIPT receivers. Here, the rate-splitting multiple access (RSMA) \cite{RSMAtrend} with the superiority to space-division multiple-access (SDMA) and non-orthogonal multiple access (NOMA) is adopted for supporting the multiple vehicles. We aim at maximizing the sum rate of vehicles by jointly optimizing the trajectory and phase shift of AIRS, transmit power and rate allocation for RSMA along with the PS ratio, for which the algorithmic solution is developed by using alternating optimization (AO) method coupled with successive convex approximation (SCA) method. The performance superiority of the proposed algorithm is verified via simulations compared to the conventional partial optimizations. To the best of our knowledge, since the RSMA-based vehicular communications with the aid of UAV or IRS are at the beginning stage, this work can provide insight and roadmap for future intelligent vehicular environments.
\vspace{-0.2cm}

\section{System model}
\begin{figure}[t]
    \centering
    \includegraphics[width=6.7cm]{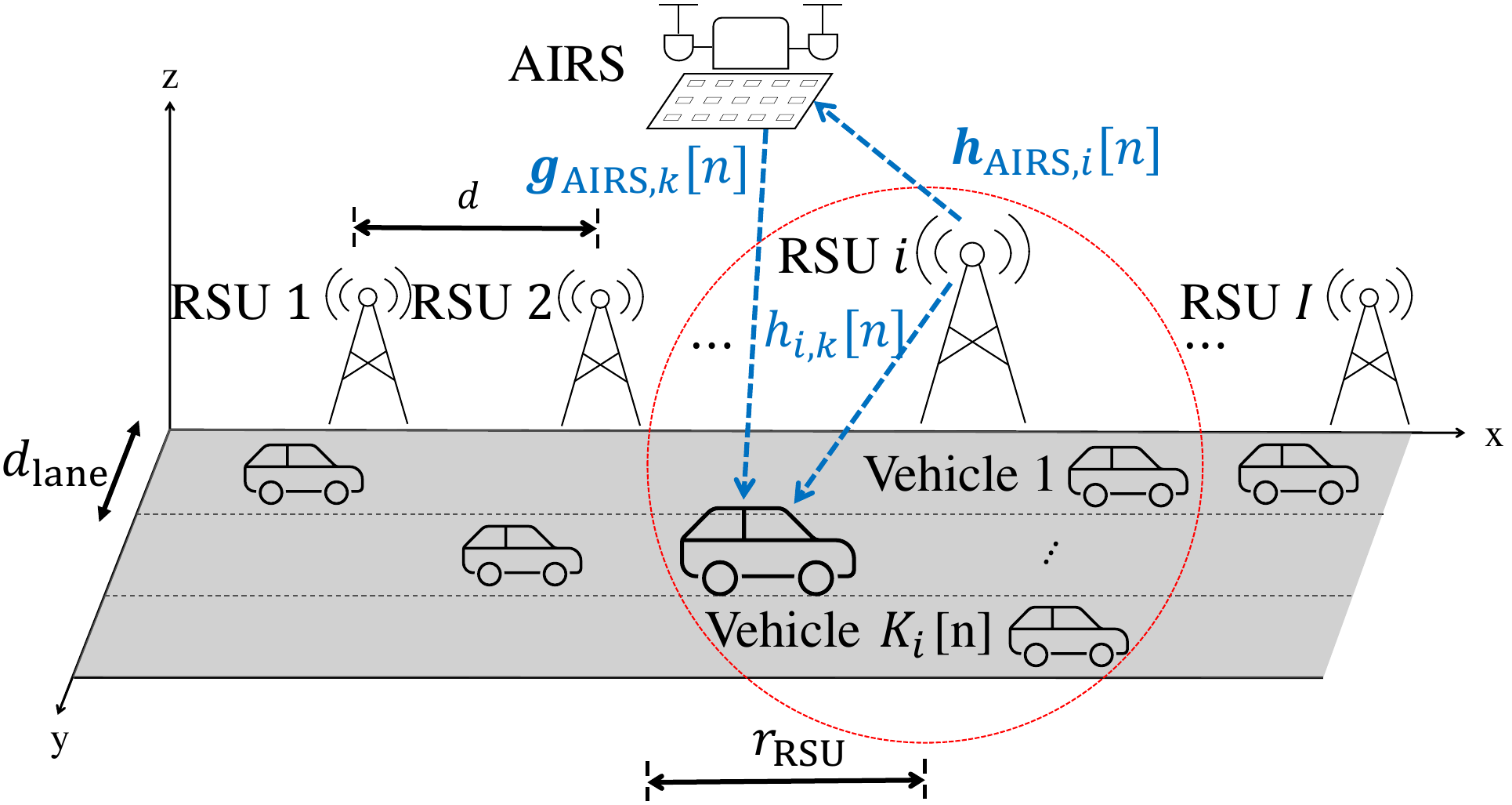}
    \caption{System model of AIRS-assisted vehicular networks with rate-splitting \\ SWIPT receivers}
\end{figure}
In this correspondence, we consider an AIRS-assisted vehicular networks as shown in Fig. 1, consisting of $K$ single-antenna vehicles, $I$ single-antenna road side units (RSUs) and an AIRS with $M$ IRS elements. Each RSU $i$ is located at $\boldsymbol{p}_{i}=(x_i, y_i)=(r_{\text{RSU}}+(i-1)d_{\text{RSU}}, 0)$ and manages $K_{i}(t)$ vehicles at time $t$, where the communications of RSUs are assumed in different orthogonal bands without interference. The coverage of each RSUs is $r_{\text{RSU}}$ m and the distance between adjacent RSUs is $d_{\text{RSU}}$ m. The sets of $I$ RSUs and $M$ elements of AIRS are defined as $\mathcal{I}\!=\!\{1, ..., I\}$ and $\mathcal{M}\!=\!\{1, ..., M\}$, respectively. For the support of $K$ battery-limited vehicles, the rate-splitting SWIPT receivers are considered for vehicles based on PS method. Also, each vehicle is assumed to arrive at the coverage edge of the first RSU at different time $t_{k}\in\{t_{1}, ..., t_{K}\}$, and each vehicle is assumed to operate in one of the $J$ lanes without changing lanes, where $\mathcal{J}\!=\!\{1, ..., J\}$. For simplicity, the velocity of vehicles in the same lane is assumed to be fixed as $v_j\in\{v_1, ..., v_J\}$. \\
\indent For tractability, we adopt the time discretization technique \cite{SCA}, and the entire communication time $T$ is divided into $N$ equal time slots, with the slot duration of $\delta$, i.e., $T\!=\!N\delta$. Accordingly, the location of the AIRS can be represented as $\boldsymbol{q}[n]\!=\![x[n], y[n], H_U]^T$, for $1\!\leq\!n\!\leq\!N$, and the set of vehicles managed by the RSU $i$ at time slot $n$ is expressed as $\mathcal{K}_{i}[n]\!=\!\{1, ..., K_{i}[n]\}$. Moreover, due to the aviation regulations of countries, the initial and the final AIRS's locations are predetermined as $\boldsymbol{q}[1]\!=\!\boldsymbol{q}_0$ and $\boldsymbol{q}[N]\!=\!\boldsymbol{q}_f$, respectively, and the maximum speed constraint of AIRS is given as $\|\boldsymbol{q}[n]-\boldsymbol{q}[n-1]\|\!\leq\! V_{\text{max}}\delta $, where $V_{\text{max}}$ indicates the maximum speed of the AIRS in m/s \cite{UAVtrajectory}. 
The vehicle $k$'s position in the $j$th lane managed by the closest RSU $i$ at the time slot $n$ is similarly represented as $\boldsymbol{p}_{i,k}[n]\!=\!(x_{i,k}[n], y_{i,k}[n])\!=\!((n\delta-t_{k}) v_j, (j-1)d_{\text{lane}})$, where $d_\text{{lane}}$ indicates the distance between neighboring lanes. \\
\indent By following \cite{VANET, mainRef}, we assume that the channel links between two nodes of interest are dominated by the line-of-sight (LoS) channel, and Doppler effects can be perfectly compensated at transceivers. The LoS channel links between the RSU $i$-to-vehicle $k$, the RSU $i$-to-AIRS and AIRS-to-vehicle $k$ at time slot $n$ can be then written as $h_{i,k}[n]\!=\!\sqrt{\smash[b]{h_0d^{-2}_{i,k}[n]}}$, $\boldsymbol{h}_{i,\text{AIRS}}[n]\!=\!\sqrt{\smash[b]{h_1d^{-2}_{i,\text{AIRS}}[n]}} [1 \; ... \; e^{-j(2\pi \slash \lambda)(M-1)d_{M}\cos\phi_i[n]}]^T$, $\boldsymbol{g}_{\text{AIRS},k}[n]\!=\!\sqrt{\smash[b]{h_1d_{\text{AIRS},k}^{-2}[n]}} [1 \; ... \; e^{-j(2\pi \slash \lambda)(M-1)d_{M}\cos\phi_k[n]}]^T$, respectively, where $h_0$ and $h_1$ denote the channel gain at the reference distance of $1$ m in ground-to-ground and air-to-ground, $d_{A,B}$ and $d_M$ are the distance between $A$ and $B$ and between two adjacent elements at AIRS, $\lambda$ is the carrier wavelength, and $\phi_i[n]$ and $\phi_k[n]$ denote the signal's angle-of-arrival from RSU $i$ to AIRS and the angle-of-departure from AIRS to the vehicle $k$ at the time slot $n$, respectively. The channel state information (CSI) of all links is assumed to be perfectly available at RSU. \\
\indent Here, we adopt the downlink RSMA technique \cite{RSMAtrend} to split into the multiple sub-messages and to allocate the different powers to these sub-messages for the time and spectral efficiency. To this end, we denote the transmit power allocation and data stream at each RSU $i$ for vehicles as $\boldsymbol{P}_i[n]\!=\!\{p_{i,0}[n], p_{i,1}[n], ..., p_{i, K_{i}[n]}[n]\}\!=\!\{p_{i,0}[n],\{p_{i,k}[n]\}_{k \in \mathcal{K}_i[n]}\}$, and $\boldsymbol{S}_i[n]\!=\!\{s_{i,0}[n], s_{i,1}[n], ..., s_{i, K_{i}[n]}[n]\}$, where the data streams $\boldsymbol{S}_i[n]$ consist of the common message $s_{i,0}[n]$ and private message $\{s_{i,k}[n]\}_{k \in \mathcal{K}_i[n]}$ \cite{RSMAtrend}. Under these assumptions, the received signal at the vehicle $k$ can be expressed as 
\begin{align}
\label{eq:received_signal}
    y_{i,k}[n] = h_{i,k}^{\text{eff}}[n]\left(\sum\nolimits_{j=0}^{K_{i}[n]} \!\sqrt{p_{i,j}[n]}s_{i,j}[n]\right) + n_{c}, 
\end{align}
where $n_{c}$ is the additive white Gaussian noise (AWGN) with $\mathcal{CN}(0, \sigma_{i,k}^2)$, and the overall effective channel of RSU $i$-AIRS-vehicle $k$ link can be expressed as
\begin{align}
    \label{eq:effective_channel}
        h_{i,k}^{\text{eff}}[n] &= h_{{i,k}}[n] + \boldsymbol{g}_{\text{AIRS},k}[n]^{H} \boldsymbol{\Theta}[n] \boldsymbol{h}_{i,\text{AIRS}}[n],
\end{align}
with $\boldsymbol{\Theta}[n]=\text{diag}[\gamma_1e^{j\theta_1[n]}, ..., \gamma_Me^{j\theta_M[n]}]$ denoting the diagonal phase shift matrix and $\theta_m[n] \in [0,2\pi)$ being the phase shifter of the $m$th element for the AIRS at time slot $n$. Without loss of generality, the reflection coefficients are assumed to be $\gamma_m\!=\!1, \; \forall m \in \mathcal{M}$ \cite{IRSdesign}. \\
\indent In the PS structure for SWIPT operation, the received signal at vehicle $k$ is divided into ID module and EH module by allocating PS ratio $\rho \in [0, 1]$, from which it can be represented as $y^{\text{ID}}_{i,k}[n]\!=\!\sqrt{\rho}y_{i,k}[n]+v_{i,k}$ and $y^{\text{EH}}_{i,k}[n]\!=\!\sqrt{1 -\rho}y_{i,k}[n]$, where $v_{i,k} \sim \mathcal{CN}(0, \epsilon_{i,k}^2)$ is the additive circuit noise at the ID module. In addition, to implement RSMA technique \cite{RSMAtrend}, the successive interference cancellation (SIC) procedure is used at the ID module to subtract the common stream, and then the private data stream is decoded by treating the remaining streams as noise. Consequently, the instantaneous achievable rates for common and private streams at vehicle $k$ can be calculated as
\begin{subequations}
\label{eq:rate_PS}
\begin{align}
    &R_{c,i,k}[n]\!=\!\log_2\!{\left(\!1\!+\!\frac{\rho(p_{i,0}[n]|h_{i,k}^{\text{eff}}[n]|^2)}{\rho\!\left(\sum_{j=1}^{K_{i}[n]}{p_{i,j}[n]|h_{i,k}^{\text{eff}}[n]|^2\!+\!\sigma_{i,k}^2}\right)\!+\!\epsilon_{i,k}^2}\!\right)\!}, \\
    &R_{i,k}[n]\!=\!\log_2\!{\left(\!1\!+\!\frac{\rho(p_{i,k}[n]|h_{i,k}^{\text{eff}}[n]|^2)}{\rho\!\left(\sum_{j=1, j\neq k}^{K_{i}[n]}{p_{i,j}[n]|h_{i,k}^{\text{eff}}[n]|^2\!+\!\sigma_{i,k}^2}\right)\!+\!\epsilon_{i,k}^2}\!\right)\!},
\end{align}
\end{subequations}
respectively. Note that since all users need to be able to decode the common data, the common message rate $R_{c,i,k}[n]$ needs to be larger than or equal to the minimum achievable rate of the users, that is, $R_{c.i.k}[n]\!\geq\!R_{c,i}[n]\!=\!\text{min}\{R_{c,i,1}[n], ..., R_{c,i,K_{i}[n]}[n]\}$. In addition, we define $C_{i,k}[n]$ as the portion of $R_{c,i}[n]$ transmitting the common parts of user $k$'s message that satisfies $\sum_{k=1}^{K_{i}[n]}{C_{i,k}[n]}\!=\!R_{c,i}[n]$. Consequently, the achievable rate of vehicle $k$ at the time slot $n$ is expressed as $R_{i,k}^{\text{tot}}[n]\!=\!C_{i,k}[n]\!+\!R_{i,k}[n]$. At the EH module, the harvested energy at the time slot $n$ can be expressed as $Q_{i,k}[n]\!=\!\zeta(1\!-\!\rho)(\sum_{j=0}^{K_{i}[n]}p_{i,j}[n]|h_{i,k}^{\text{eff}}[n]|^2)$, where $\zeta\!\in\! (0, 1]$ denotes the energy conversion efficiency of EH module of vehicles. For the EH operation, we have the requirements for the minimum harvested energy constraint as $\sum_{i=1}^I\sum_{n=1}^N Q_{i,k}[n] \geq E_{k}^{th}$, where $E_{k}^{th}$ is the threshold of harvested energy.
\hspace{-0.2cm}

\section{Problem Formulation and Proposed Algorithms}
In this section, we formulate the optimization problem to maximize the sum rate of the AIRS-assisted vehicular networks. To this end, we formulate the sum-rate maximization problem as \vspace{-0.3cm}

\begin{subequations}
\label{eq:problem_formulation}
    \begin{eqnarray}
    &&\max_{\boldsymbol{q},\boldsymbol{p},\boldsymbol{C},\boldsymbol{\theta},0\leq\rho\leq1} {\sum_{n=1}^N{R_{\text{tot}}}[n]=\sum_{n=1}^N{\sum_{i=1}^I{\sum_{k=1}^{K_{i}[n]}{R_{i,k}^{\text{tot}}[n]}}}}\label{eq:problem_formulation_objective_function}\qquad\\
    &&\qquad \text{s.t.} \quad \sum_{i=1}^{I}{{\sum_{n=1}^{N}{Q_{i,k}[n]}}\geq E_{k}^{th}}, \quad \forall i,k,\qquad\qquad \label{eq:HV_const}\qquad \\
    &&\qquad\qquad\; \sum\nolimits_{k=0}^{K_{i}[n]}p_{i,k}[n] \leq P_{i}^{\text{max}}, \quad \forall i,n, \label{eq:TP_const}\qquad \\
    &&\qquad\qquad\; \sum\nolimits_{k=1}^{K_{i}[n]}{C_{i,k}[n]} \leq R_{c,i,k}[n], \;\;\; \forall i,k,n, \label{eq:RA_const}\qquad \\
    &&\qquad\qquad\; \boldsymbol{q}[1] = \boldsymbol{q}_0, \;\; \boldsymbol{q}[N] =\boldsymbol{q}_f, \label{eq:IF_AIRS_const}\qquad \\
    &&\qquad\qquad\; \|\boldsymbol{q}[n]-\boldsymbol{q}[n-1]\| \leq V_{\text{max}}\delta, \quad \forall n, \label{eq:MV_AIRS_const}\qquad \\
    &&\qquad\qquad\; \theta_m[n] \in [0, 2\pi), \quad \forall n,m, \label{eq:PH_const}\qquad
    \end{eqnarray}
\end{subequations} where $\boldsymbol{q}\!=\!\{\boldsymbol{q}[n]\}_{\forall n}$, $\boldsymbol{p}\!=\!\{p_{i,k}[n]\!\geq\!0\}_{\forall i,k,n}$, $\boldsymbol{C}\!=\!\{C_{i,k}[n]\!\geq\!0\}_{\forall i,k,n}$, $\boldsymbol{\theta}\!=\!\{\theta_m[n]\}_{\forall m,n}$, and $P_{i}^{\text{max}}$ is the maximum transmit power constraint of RSU $i$. Due to the non-convex objective function (\ref{eq:problem_formulation_objective_function}) and non-convex constraints (\ref{eq:HV_const}), (\ref{eq:RA_const}), the problem (\ref{eq:problem_formulation}) is non-convex. To solve (\ref{eq:problem_formulation}), we propose the AO-based algorithm with SCA method, which alternately updates the subsets of variables, while keeping the others fixed, and iteratively approximates the non-convex functions with a sequence of convex surrogate functions. The details of algorithm are provided in the following.
\vspace{-0.2cm}

\subsection{Joint Phase Shifter and Trajectory Optimization}
For any given $\boldsymbol{p},\boldsymbol{C}$ and $\rho$, the joint optimization on the AIRS's phase shifter $\boldsymbol{\theta}$ and trajectory $\boldsymbol{q}$ can be reformulated from (\ref{eq:problem_formulation}) as \vspace{-0.3cm}
\begin{align}
\label{eq:sub_opt_prob1}
        \max_{\boldsymbol{q},\boldsymbol{\theta}} \quad  \sum\nolimits_{n=1}^N{R_{\text{tot}}}[n] \quad \text{s.t.} \quad (\text{\ref{eq:HV_const}}), (\text{\ref{eq:RA_const}}), (\text{\ref{eq:IF_AIRS_const}})-(\text{\ref{eq:PH_const}}).
\end{align}
To cope with the non-convexity of (\ref{eq:sub_opt_prob1}), we employ the two-step approach, where the trajectory of AIRS is obtained after optimizing the phase shifter matrix. \\
\indent Specifically, according to (\ref{eq:effective_channel}), the optimal $\boldsymbol{\theta}$ in (\ref{eq:sub_opt_prob1}) needs to maximize $\sqrt{p_{i,k}[n]}h_{i,k}^{\text{eff}}[n]$, which is obtained when the exponent values of $h_{i,k}^{\text{eff}}[n]$ is substituted by zero. Therefore, the $m$th element of the optimal phase shifter $\check{\theta}_m[n]$ at time slot $n$ can be calculated as 
\begin{equation}
\label{eq:opt_PH}
    \check{\theta}_m[n] = \frac{2\pi}{\lambda}d_M(m-1)(\cos{\phi_i[n]}-\cos{\phi_k[n]}).
\end{equation}
With the fixed phase shifter $\boldsymbol{\check{\theta}}=\{\check{\theta}_m[n]\}_{\forall m,n}$ in (\ref{eq:opt_PH}) and slack variables $\boldsymbol{u}\!=\!\{u_{i,k}[n]\}_{\forall i,k,n}$, $\boldsymbol{v}\!=\!\{v_i[n]\}$ $_{\forall i,n}$, $\boldsymbol{w}\!=\!\{w_{i,k}[n]\}_{\forall i,k,n}$ and $\boldsymbol{o}\!=\!\{o_i[n]\}_{\forall i,n}$ and by applying the first-order Taylor approximation \cite{SCA}, the optimization problem of the AIRS trajectory $\boldsymbol{q}$ can be written as 
\vspace{-0.2cm}
\begingroup
\allowdisplaybreaks
\begin{subequations}
\label{eq:sub_opt_prob1_25}
    \begin{align}
        \max_{\boldsymbol{q},\boldsymbol{u},\boldsymbol{w},\boldsymbol{v},\boldsymbol{o}} \quad & {\sum\nolimits_{n=1}^N{R_{\text{tot}}^{\text{lb}}[n]}} \\
        \text{s.t.} \qquad & \sum\nolimits_{i=1}^I\sum\nolimits_{n=1}^NQ_{i,k}^{\text{lb}}[n]\geq E_{k}^{th}, \quad \forall i,k, \label{eq:sub_opt_prob1_25b} \\
        & \sum\nolimits_{k=1}^{K_{i}[n]}{C_{i,k}[n]} \leq R_{c,i,k}^{\text{lb}}[n], \quad \forall i,k,n, \label{eq:sub_opt_prob1_25c} \\
        & u_{i,k}[n] \geq d_{\text{AIRS},k}^2[n], v_i[n] \geq d_{i,\text{AIRS}}^2[n], \label{slack_variabl_uv} \\
        & w_{i,k}[n]\leq d_{\text{AIRS},k}^{\text{lb}}[n], o_i[n] \leq d_{i,\text{AIRS}}^{\text{lb}}[n], \label{slack_variabl_wo} \\
        & (\text{\ref{eq:IF_AIRS_const}}), (\text{\ref{eq:MV_AIRS_const}}), \label{eq:sub_opt_prob1_25h}
    \end{align}
\end{subequations}
\endgroup
where the optimal overall effective channel of RSU $i$-AIRS-vehicle $k$ link is expressed as $\; \check{h}_{i,k}^{\text{eff}}[n]\; =\; \sqrt{h_0}\slash d_{i,k}[n] \;+\; h_1M \slash (d_{i,\text{AIRS}}[n]d_{\text{AIRS},k}[n])$, ${R}^{\text{lb}}_{\text{tot}}[n]=\sum_{i=1}^I{\sum_{k=1}^{K_{i}[n]}{{R}^{\text{lb}}_{i,k}[n]}} + \sum_{i=1}^I{\text{min}\{{R}^{\text{lb}}_{c,i,1}[n], ...,{R}^{\text{lb}}_{c,i,K_i[n]}[n]\}}$, $d_{\text{AIRS},k}^{\text{lb}}[n]=(x^{(l)}[n]\!-\!x_{i,k}[n])^2\!+\!(y^{(l)}[n]\!-\!y_{i,k}[n])^2\!+\!H_U^2\!+\!2(x^{(l)}[n]\!-\!x_{i,k}[n])(x[n]\!-\!x^{(l)}[n])\!+\!2(y^{(l)}[n]\!-\!y_{i,k}[n])(y[n]\!-\!y^{(l)}[n])$ and $d_{i,\text{AIRS}}^{\text{lb}}[n]\!=\!(x^{(l)}[n]\!-\!x_{i}[n])^2\!+\!(y^{(l)}[n]\!-\!y_{i}[n])^2\!+\!H_U^2\!+\!2(x^{(l)}[n]\!-\!x_{i}[n])(x[n]\!-\!x^{(l)}[n])\!+\!2(y^{(l)}[n]\!-\!y_{i}[n])(y[n]\!-\!y^{(l)}[n])$, which are updated in the $l$th iteration. Furthermore, the slack variables are used to approximate $|\check{h}_{i,k}^{\text{eff}}[n]|^2$ by the lower and upper bounds as 
\begin{subequations}
\label{eq:sl_variables}
    \begin{align}
    &|\check{h}_{i,k}^{\text{eff}}[n]|^2\!\geq\! A_{i,k}[n]\!+\!\frac{B_{i,k}[n]}{u_{i,k}^{1/2}[n]v_i^{1/2}[n]}\!+\!\frac{C}{u_{i,k}[n]v_i[n]}\!\triangleq\!h_{i,k}^{\text{eff}, \dag}[n], \\
    &|\check{h}_{i,k}^{\text{eff}}[n]|^2\!\leq\! A_{i,k}[n]\!+\!\frac{B_{i,k}[n]}{w_{i,k}^{1/2}[n]o_i^{1/2}[n]}\!+\!\frac{C}{w_{i,k}[n]o_i[n]}\!\triangleq\!h_{i,k}^{\text{eff}, \ddag}[n],
    \end{align}
\end{subequations}
with $A_{i,k}[n]={h_0}\slash{d_{i,k}^2[n]}$, $B_{i,k}[n]={2h_1^{3/2}M}\slash{d_{i,k}[n]} \;\, \text{and} \;\, C \\
=\;h_12M^2$. Also $R_{c,i,k}^{\text{lb}}[n]$ and $R_{i,k}^{\text{lb}}$ are obtained by first-order Taylor approximation \cite{SCA} as
\vspace{-0.2cm}
\begin{subequations}
\begin{align}
    &R_{c,i,k}^{\text{lb}}[n]=\log_2{(X_{i,k,\dag}^{(l)}[n])}+\frac{Y_{i,k,\dag}^{(l)}[n]}{X_{i,k,\dag}^{(l)}[n]\ln2}(u_{i,k}[n]-u_{i,k}^{(l)}[n]) + \nonumber \\
    &\frac{Z_{i,k,\dag}^{(l)}[n]}{X_{i,k,\dag}^{(l)}[n]\ln2}(v_i[n]-v_i^{(l)}[n]) - \log_2{(X_{i,k,\ddag}^{(l)}[n])} - \frac{Y_{i,k,\ddag}^{(l)}[n]}{X_{i,k,\ddag}^{(l)}[n]\!\ln2} \nonumber \\
    &(w_{i,k}[n]\!-\!w_{i,k}^{(l)}[n])\!-\!\frac{Z_{i,k,\ddag}^{(l)}[n]}{X_{i,k,\ddag}^{(l)}[n]\!\ln2}\!(o_i[n]\!-\!o_i^{(l)}[n]), \\
    &R_{i,k}^{\text{lb}}[n]=\log_2{(F_{i,k,\dag}^{(l)}[n])}+\frac{G_{i,k,\dag}^{(l)}[n]}{F_{i,k,\dag}^{(l)}[n]\ln2}(u_{i,k}[n]-u_{i,k}^{(l)}[n]) + \nonumber \\
    &\frac{I_{i,k,\dag}^{(l)}[n]}{F_{i,k,\dag}^{(l)}[n]\ln2}(v_i[n]-v_i^{(l)}[n]) - \log_2{(F_{i,k,\ddag}^{(l)}[n])} - \frac{G_{i,k,\ddag}^{(l)}[n]}{F_{i,k,\ddag}^{(l)}[n]\!\ln2} \nonumber \\
    &(w_{i,k}[n]\!-\!w_{i,k}^{(l)}[n])\!-\!\frac{I_{i,k,\ddag}^{(l)}[n]}{F_{i,k,\ddag}^{(l)}[n]\!\ln2}\!(o_i[n]\!-\!o_i^{(l)}[n]),
\end{align}
\end{subequations}
where the updated values of slack variables in the $l$th iteration are written as 
$X_{i,k,*}^{(l)}[n]\!=\!\rho(\sum_{j=\alpha}^{K_{i}[n]}p_{i,j}[n](A_{i,k}[n]\!+\!{B_{i,k}[n]} \slash {(f_{i,k}^{(l)}[n]g_i^{(l)}[n])^{1/2}}\!+\!{C} \slash {(f_{i,k}^{(l)}[n]g_i^{(l)}[n])})\!+\!\sigma_{i,k}^2)\!+\!\epsilon_{i,k}^2$, $Y_{i,k,*}^{(l)}[n]\!=\!-\rho\sum_{j=\alpha}^{K_{i}[n]}p_{i,j}[n]({B_{i,k}[n]} \slash {(2(f_{i,k}^{(l)}[n])^{3/2}(g_i^{(l)}[n])^{1/2})}\!+\!{C}\slash{((f_{i,k}^{(l)}[n])^2g_i^{(l)}[n]}))$, $Z_{i,k,*}^{(l)}[n]\!=\!-\rho\sum_{j=\alpha}^{K_{i}[n]}p_{i,j}[n]({B_{i,k}[n]}\slash{(2(f_{i,k}^{(l)}[n])^{1/2}(g_i^{(l)}[n])^{3/2})}\!+\!{C}\slash{(u_{i,k}^{(l)}[n](v_i^{(l)}[n])^2}))$, $F_{i,k,*}^{(l)}[n] \\ \!=\!\rho(\sum_{j=\beta}^{K_{i}[n]}p_{i,j}[n](A_{i,k}[n]\!+\!{B_{i,k}[n]} \slash {(f_{i,k}^{(l)}[n]g_i^{(l)}[n])^{1/2}}\!+\!{C} \slash {(f_{i,k}^{(l)}[n]g_i^{(l)}[n]}))\!+\sigma_{i,k}^2)\!+\epsilon_{i,k}^2$, $G_{i,k,*}^{(l)}[n]=-\rho\sum_{j=\beta}^{K_{i}[n]}p_{i,j}[n]\\({B_{i,k}[n]} \slash {(2(f_{i,k}^{(l)}[n])^{3/2}(g_i^{(l)}[n])^{1/2})}\!+\!{C}\slash{((f_{i,k}^{(l)}[n])^2g_i^{(l)}[n]}))$ and  $I_{i,k,*}^{(l)}[n] = -\rho\sum_{j=\beta}^{K_{i}[n]}p_{i,j}[n]({B_{i,k}[n]} \slash {(2(f_{i,k}^{(l)}[n])^{1/2}}$ 
$(g_i^{(l)}[n])^{3/2})+{C}\slash{(u_{i,k}^{(l)}[n](v_i^{(l)}[n])^2}))$, which satisfies $*\!\in\!\{\dag, \ddag\}$, $\alpha\!=\!0$, $f\!=\!u$, $g\!=\!v$ and $\beta\!=\!1$ if $*\!=\!\dag$, otherwise $\alpha\!=\!1$, $f\!=\!w$, $g\!=\!o$ and $\beta\!=\!1,j\neq k$. Moreover, $Q_{i,k}^{\text{lb}}[n]$ is obtained similarly as $Q_{i,k}^{\text{lb}}[n]\;=\; J_{i,k}^{(l)}[n]\;+\;K_{i,k}^{(l)}[n](u_{i,k}[n]\;-\\\!u_{i,k}^{(l)}[n])+L_{i,k}^{(l)}[n](v_i[n]-v_i^{(l)}[n])$, where $J_{i,k}^{(l)}[n] \!=\! \zeta(1\!-\!\rho)\sum_{j=0}^{K_{i}[n]}p_{i,j}[n](A_{i,k}[n]\!+\!B_{i,k}[n]\slash(u_{i,k}^{(l)}[n]v_i^{(l)}[n])^{1/2}\;+\;C\slash(u_{i,k}^{(l)}[n]v_i^{(l)}[n]))$, $K_{i,k}^{(l)}[n]\;=\;\zeta(1\!-\!\rho)Y_{i,k,\dag}^{(l)}[n] \slash \rho$, $L_{i,k}^{(l)}[n]\!=\!-\zeta\\(1\!-\!\rho)Z_{i,k,\dag}^{(l)}[n] \slash \rho$. Finally, the problem (\ref{eq:sub_opt_prob1_25}) is convex that can be solved by a standard optimization solver, e.g., CVX \cite{CVX}.
\vspace{-0.2cm}

\subsection{Joint Power and Rate Allocation Optimization}
Given AIRS's trajectory $\boldsymbol{q}$, phase shifter $\boldsymbol{\theta}$ and the PS ratio $\rho$, the problem (\ref{eq:problem_formulation}) can be rewritten as \vspace{-0.1cm}
\begin{align}
    \label{eq:sub_opt_prob2}
        \max_{\boldsymbol{C},\boldsymbol{p}} \quad \sum\nolimits_{n=1}^N{R_{\text{tot}}[n]} \quad \text{s.t.} \quad (\text{\ref{eq:HV_const}})-(\text{\ref{eq:RA_const}}).
\end{align}
Since the problem (\ref{eq:sub_opt_prob2}) is non-convex due to non-convexity of (\ref{eq:problem_formulation_objective_function}) and (\ref{eq:RA_const}), we apply the difference-of-convex (DC) approximation. Specifically, the instantaneous rate for common and private messages $R_{c,i,k}[n]$ and $R_{i,k}[n]$ between vehicle $k$ and RSU $i$ can be lower-bounded as
\begingroup
\allowdisplaybreaks
\begin{subequations}
\begin{align}
    \label{eq:rate_DC1_approximation}
    &R_{c,i,k}^{\text{dc1}}[n]\!=\!\log_2{\Big(\rho\Big(\sum\nolimits_{j=0}^{K_{i}[n]}p_{i,j}[n]|h_{i,k}^{\text{eff}}[n]|^2\!+\!\sigma_{i,k}^2\Big)\!+\!\epsilon_{i,k}^2\Big)}\nonumber \\
    &- \log_2{\Big(\rho\Big(\sum\nolimits_{j=1}^{K_{i}[n]}p^{(l)}_{i,j}[n]|h_{i,k}^{\text{eff}}[n]|^2\!+\!\sigma_{i,k}^2\Big)\!+\!\epsilon_{i,k}^2\Big)}\nonumber \\ 
    &-\frac{\rho(|h_{i,k}^{\text{eff}}[n]|^2+\sigma_{i,k}^2)(\sum_{j=1}^{K_{i}[n]}p_{i,j}[n]-p^{(l)}_{i,j}[n])}{\ln{2}\!(\rho(\sum_{j=1}^{K_{i}[n]}p^{(l)}_{i,j}[n]|h_{i,k}^{\text{eff}}[n]|^2+\sigma_{i,k}^2)+\epsilon_{i,k}^2)}, \\
    &R_{i,k}^{\text{dc1}}[n] \!=\! \log_2{\Big(\rho\Big(\sum\nolimits_{j=1}^{K_{i}[n]}p_{i,j}[n]|h_{i,k}^{\text{eff}}[n]|^2\!+\!\sigma_{i,k}^2\Big)\!+\!\epsilon_{i,k}^2\Big)}\nonumber \\
    &- \log_2{\Big(\rho\Big(\sum\nolimits_{j=1,j \neq k}^{K_{i}[n]}p^{(l)}_{i,j}[n]|h_{i,k}^{\text{eff}}[n]|^2\!+\!\sigma_{i,k}^2\Big)\!+\!\epsilon_{i,k}^2\Big)}\nonumber \\ 
    &-\frac{\rho(|h_{i,k}^{\text{eff}}[n]|^2+\sigma_{i,k}^2)(\sum_{j=1,j \neq k}^{K_{i}[n]}p_{i,j}[n]-p^{(l)}_{i,j}[n])}{\ln{2}(\rho(\sum_{j=1, j \neq k}^{K_{i}[n]}p^{(l)}_{i,j}[n]|h_{i,k}^{\text{eff}}[n]|^2+\sigma_{i,k}^2)+\epsilon_{i,k}^2)},
\end{align}    
\end{subequations}
\endgroup
where the updated values in the $l$th iteration can be expressed as $\boldsymbol{p^{(l)}}\!=\!\{p^{(l)}_{i,k}[n]\}_{\forall{i,k,n}}$, and the objective function is given as $R_{\text{tot}}^{\text{dc1}}[n]\!=\!\sum_{i=1}^{I}{\text{min}\{{R}^{\text{dc1}}_{c,i,1}[n],...,{R}^{\text{dc1}}_{c,i,K_i[n]}[n]\}}\!+\!\sum_{i=1}^{I}{\sum_{k=1}^{K_{i}[n]}{R_{i,k}^{\text{dc1}}[n]}}$. Therefore, the optimization problem (\ref{eq:sub_opt_prob2}) is expressed as \vspace{-0.2cm}
\begingroup
\allowdisplaybreaks
\begin{subequations}
\label{eq:sub_opt_prob2_end}
    \begin{align}
        \max_{\boldsymbol{C},\boldsymbol{p}} \qquad & {\sum\nolimits_{n=1}^N{R^{\text{dc1}}_{\text{tot}}[n]}} \\
        \text{s.t.} \qquad \; & \sum\nolimits_{k=1}^{K_{i}[n]}{C_{i,k}[n]} \leq R_{c,i,k}^{\text{dc1}}[n] \quad \forall i,k,n \\
        & (\text{\ref{eq:HV_const}}), (\text{\ref{eq:TP_const}}),
    \end{align}
\end{subequations}
which is convex and whose solution can be found, e.g., by CVX \cite{CVX}.
\vspace{-0.2cm}

\subsection{PS Ratio Optimization}
Finally, we optimize the PS ratio $\rho$ by fixing the variables $\boldsymbol{p}, \boldsymbol{C}, \boldsymbol{q}$ and $\boldsymbol{\theta}$, whose problem is given as \vspace{-0.2cm}\
\begin{align}
\label{eq:sub_opt_prob3}
        \max_{0\leq\rho\leq1} \quad {\sum\nolimits_{n=1}^N{R_{\text{tot}}[n]}}  \quad \text{s.t.} \quad (\text{\ref{eq:HV_const}}), (\text{\ref{eq:RA_const}}).
\end{align} 
\endgroup
Since (\ref{eq:sub_opt_prob3}) is non-convex, by using the DC approximation, we have the approximation of (\ref{eq:problem_formulation_objective_function}) as $R_{\text{tot}}^{\text{dc2}}[n]\!=\!\sum_{i=1}^{I}{\text{min}\{{R}^{\text{dc2}}_{c,i,1}[n],...,{R}^{\text{dc2}}_{c,i,K_i[n]}[n]\}}\!+\!\sum_{i=1}^{I}{\sum_{k=1}^{K_{i}[n]}{R_{i,k}^{\text{dc2}}[n]}}$. Similar to (\ref{eq:rate_DC1_approximation}), the lower-bounded instantaneous rate for common and private messages $R_{c,i,k}^{\text{dc2}}[n]$ and $R_{i,k}^{\text{dc2}}[n]$ between vehicle $k$ and RSU $i$ can be expressed as
\begingroup
\allowdisplaybreaks
\begin{subequations}
\begin{align}
    \label{eq:rate_DC2_approximation}
    &R_{c,i,k}^{\text{dc2}}[n]\!=\!\log_2{\Big(\rho\Big(\sum\nolimits_{j=0}^{K_{i}[n]}p_{i,j}[n]|h_{i,k}^{\text{eff}}[n]|^2\!+\!\sigma_{i,k}^2\Big)\!+\!\epsilon_{i,k}^2\Big)}\nonumber \\
    &- \log_2{\Big(\rho^{(l)}\Big(\sum\nolimits_{j=1}^{K_{i}[n]}p_{i,j}[n]|h_{i,k}^{\text{eff}}[n]|^2\!+\!\sigma_{i,k}^2\Big)\!+\!\epsilon_{i,k}^2\Big)}\nonumber \\ 
    &-\frac{(\sum\nolimits_{j=1}^{K_{i}[n]}p_{i,j}[n]|h_{i,k}^{\text{eff}}[n]|^2+\sigma_{i,k}^2)(\rho-\rho^{(l)})}{\ln{2}(\rho^{(l)}(\sum_{j=1}^{K_{i}[n]}p^{(l)}_{i,j}[n]|h_{i,k}^{\text{eff}}[n]|^2+\sigma_{i,k}^2)+\epsilon_{i,k}^2)}, \\
    &R_{i,k}^{\text{dc2}}[n]\!=\!\log_2{\Big(\rho\Big(\sum\nolimits_{j=1}^{K_{i}[n]}p_{i,j}[n]|h_{i,k}^{\text{eff}}[n]|^2\!+\!\sigma_{i,k}^2\Big)\!+\!\epsilon_{i,k}^2\Big)}\nonumber \\
    &- \log_2{\Big(\rho^{(l)}\Big(\sum\nolimits_{j=1,j \neq k}^{K_{i}[n]}p_{i,j}[n]|h_{i,k}^{\text{eff}}[n]|^2\!+\!\sigma_{i,k}^2\Big)\!+\!\epsilon_{i,k}^2\Big)}\nonumber \\ 
    &-\frac{(\sum\nolimits_{j=1,j \neq k}^{K_{i}[n]}p_{i,j}[n]|h_{i,k}^{\text{eff}}[n]|^2+\sigma_{i,k}^2)(\rho-\rho^{(l)})}{\ln{2}(\rho^{(l)}(\sum_{j=1,j \neq k}^{K_{i}[n]}p^{(l)}_{i,j}[n]|h_{i,k}^{\text{eff}}[n]|^2+\!\sigma_{i,k}^2)+\!\epsilon_{i,k}^2)}.
\end{align}    
\end{subequations}
\endgroup
Accordingly, the optimization problem (\ref{eq:sub_opt_prob3}) can be expressed as \vspace{-0.1cm}
\begingroup
\allowdisplaybreaks
\begin{subequations}
\label{eq:sub_opt_prob3_end}
    \begin{align}
        \max_{\rho} \qquad &{\sum\nolimits_{n=1}^N{R_{\text{tot}}^{\text{dc2}}[n]}} \\
        \text{s.t.} \qquad \; &\sum\nolimits_{k=1}^{K_{i}[n]}{C_{i,k}[n]} \leq R_{c,i,k}^{\text{dc2}}[n], \\
        & (\text{\ref{eq:HV_const}}),
    \end{align}    
\end{subequations}
\endgroup
which is convex, and can be readily solved e.g., by CVX. \\
\indent The overall proposed algorithm for the sum-rate maximization of AIRS-assisted vehicular networks is summarized in Algorithm 1.
\begin{algorithm}
\caption{\footnotesize Proposed Algorithm} \label{alg1}
\begin{algorithmic}
    \footnotesize
    \renewcommand{\algorithmicrequire}{\textbf{Initialize:}}
    \Require
    Initialize $\boldsymbol{q}^{(0)}, \boldsymbol{p}^{(0)}, \boldsymbol{C}^{(0)}$ and $\rho^{(0)}$ by setting $l=0$, and calculate $\boldsymbol{\theta}^{(0)}$ via (\ref{eq:opt_PH}).
    \renewcommand{\algorithmicrequire}{\textbf{Repeat:}}
    \Require Until the convergence criterion is satisfied.\\
    Update $\boldsymbol{q}^{(l)}$ via (\ref{eq:sub_opt_prob1_25}), given $\boldsymbol{p}^{(l-1)}, \boldsymbol{C}^{(l-1)},\rho^{(l-1)}$, $\boldsymbol{\theta}^{(l-1)}$.\\
    Optimize $\boldsymbol{\theta}^{(l)}$ via (\ref{eq:opt_PH}). \\
    Compute $\boldsymbol{p}^{(l)}$ and $\boldsymbol{C}^{(l)}$ by (\ref{eq:sub_opt_prob2_end}), given $\rho^{(l-1)}$, $\boldsymbol{q}^{(l)}$, $\boldsymbol{\theta}^{(l)}$. \\
    Optimize $\rho^{(l)}$ by (\ref{eq:sub_opt_prob3_end}), given $\boldsymbol{q}^{(l)}, \boldsymbol{p}^{(l)}$, $\boldsymbol{C}^{(l)}$.\\
    Update $l \leftarrow l + 1$
    \renewcommand{\algorithmicensure}{\textbf{Output:}} 
    \Ensure 
    $\boldsymbol{q}^{*}, \boldsymbol{p}^{*}, \boldsymbol{C}^{*},\rho^{*}$
\end{algorithmic}
\label{alg1}
\end{algorithm}
\vspace{-0.2cm}

\section{Numerical Results}
\begin{table}[b]
    \footnotesize
        \centering
        \caption{\centering Simulation Parameters}
        \vspace{-0.25cm}
    \begin{tabular}{c c c c}
    \hline\hline
    Parameter & Value & Parameter & Value \\ [0.5ex] 
    \hline
    $r_{\text{RSU}}$ & 250 m & $h_0,h_1$ & 0,20 dB \\ 
    $d_{\text{lane}}$ & 4 m & $d_M$ & 0.05 m \\
    $\delta$ & 1 s & $q_0$ [m] & \{250, 10, 20\} \\
    $T$ & 50 s & $q_f$ [m] & \{1250, 10, 20\} \\
    $K$ & 4 & $\zeta$ & 0.97 \\
    $I$ & 3 & $\sigma^2_{i,k}$ & -70 dB \\
    $t_k$ [s] & \{0, 5, 20, 20\} & $\epsilon^2_{i,k}$ & -70 dB \\
    $M$ & 16 & $E_{k}^{th}$ & -50 dBm \\
    $J$ & 3 & $P_{i}^{\text{max}}$ & 29 dBm \\
    $v_j$ [m/s] & \{25, 27, 30\} & $V_{\text{max}}$ & 40 m/s \\ [1ex] 
    \hline\hline
    \end{tabular}
    \label{tab:my_label}
\end{table}
In this section, the performances of the proposed scheme are verified via simulations compared to the several benchmark schemes including existing techniques \cite{TOgraph, POgraph}. We consider the highway scenario, where the RSUs are spaced apart at regular interval $d_{\text{RSU}}=500$ m. By following the recent studies on AIRS-assisted communication systems \cite{mainRef}, the remaining simulation parameters are set as in Table 1. \\
\indent In Fig. 2, the convergence of proposed algorithm is verified numerically within about 50 iterations. Fig. 3. shows the optimized AIRS trajectory by the proposed algorithm (in the first row) along with the partial trajectory at every 25 s (in the second and three row) according to the time-varying positions of vehicles. The optimized trajectory of AIRS in black-dot line is designed to spend the sufficient time hovering near the RSU for communications under the AIRS mobility constraints and to move toward the next RSU except the hovering time at a speed close to $V_{\text{max}}$.

\begin{figure}[t]
    \centering
    \includegraphics[width=6.6cm]{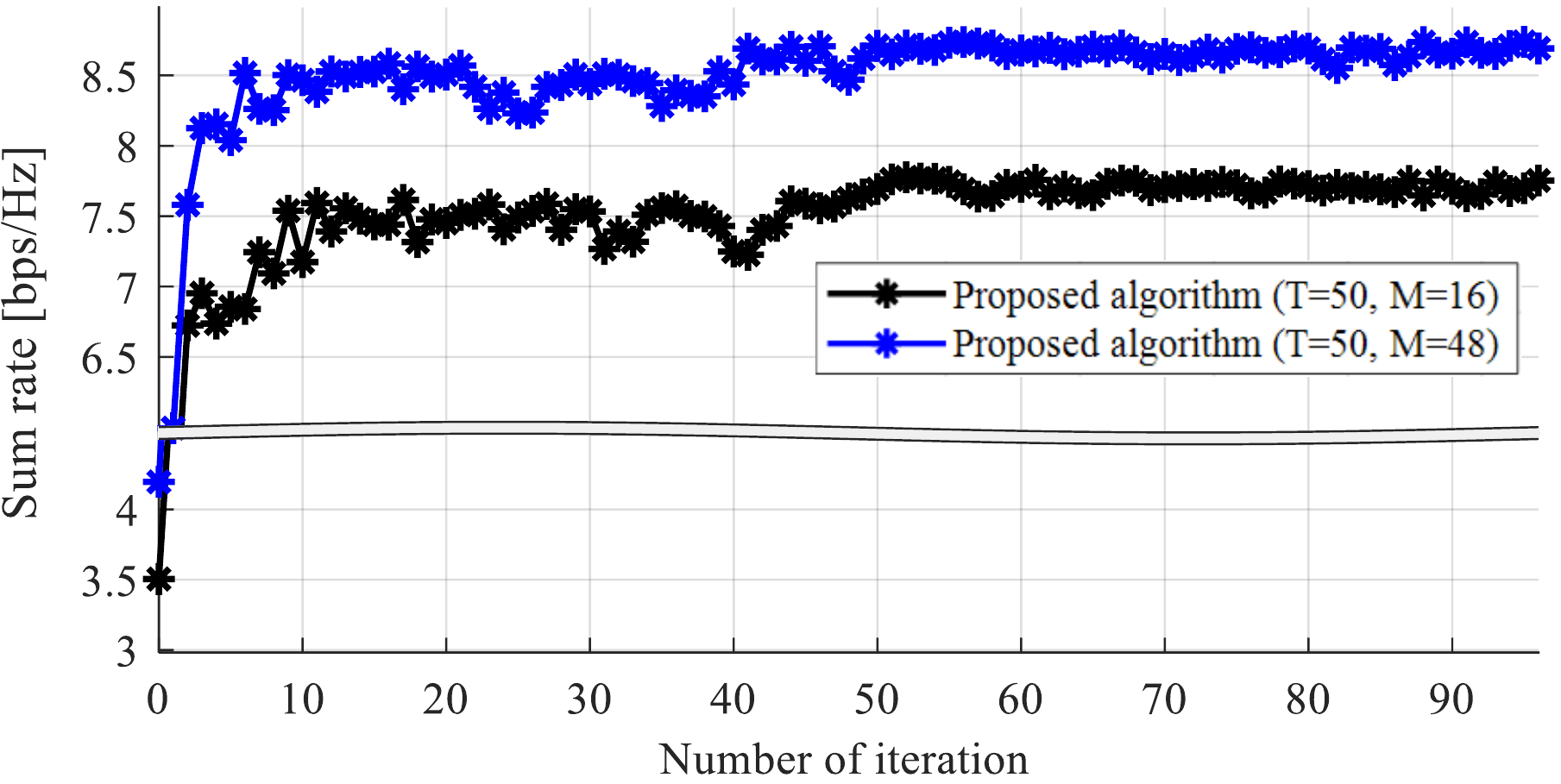}
    \caption{The convergence of proposed Algorithm 1}
\end{figure}
\begin{figure}[t]
    \centering
    \hspace{0.75cm}
    \includegraphics[width=7.15cm]{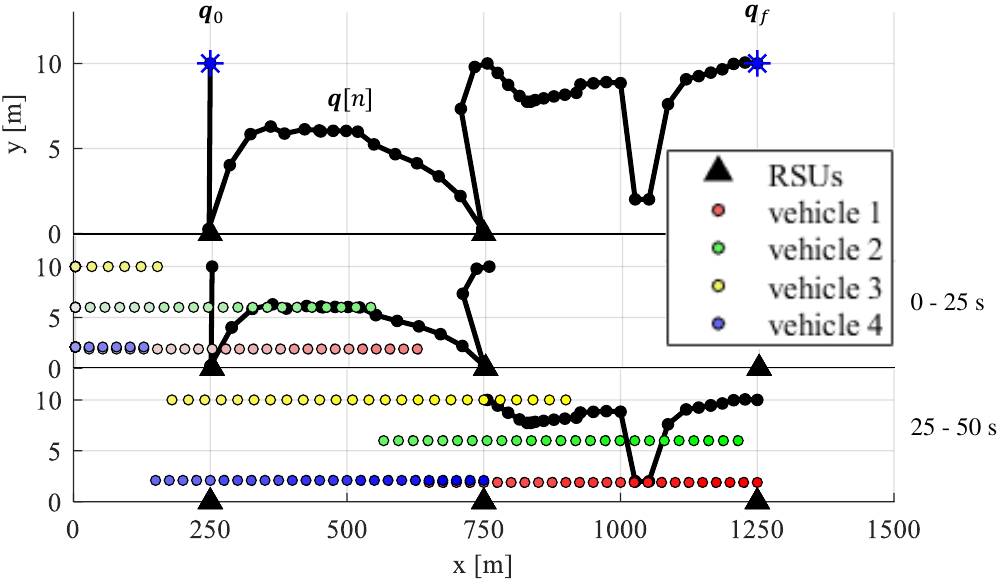}
    \caption{The AIRS's optimal trajectory of obtained by proposed Algorithm 1}
\end{figure}
\indent In Fig. 4 and Fig. 5, the sum-rate performances of proposed algorithm are illustrated for the different number $M$ of AIRS elements and the different mission time $T$, compared with reference schemes such as sole or partial optimization. It is observed that the proposed algorithm can provide the highest sum-rate performance for the different $M$ and $T$. Moreover, as the number of elements in AIRS and the communication time increase, the benefits of joint optimization become pronounced since the more number of AIRS elements can strengthen the beam alignment for the virtual LoS link, while the larger communication time can provide the more resources for the optimization.
\begin{figure}[t]
    \centering
    \includegraphics[width=6.6cm]{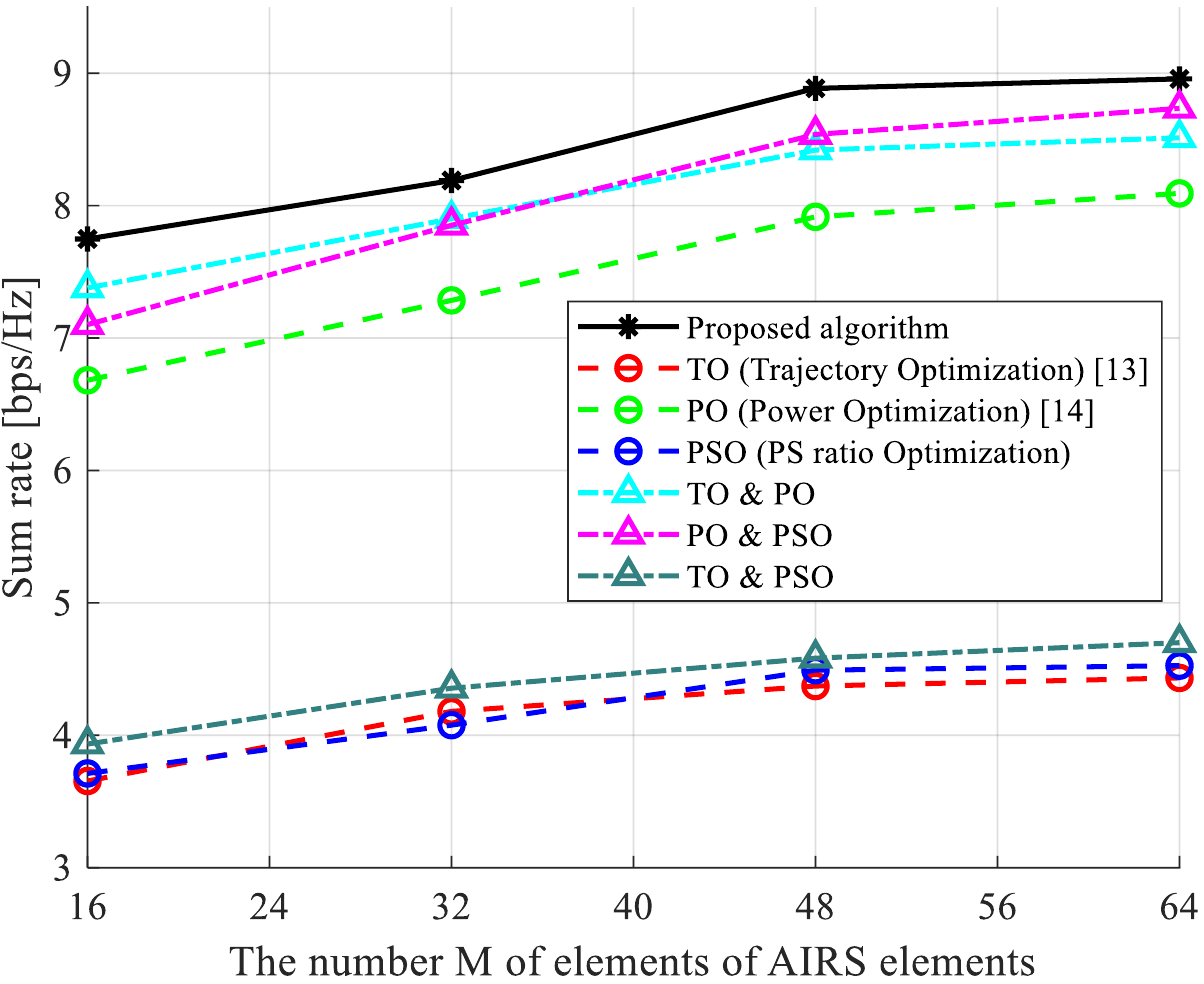}
    \caption{Sum rate versus the different number of AIRS elements}
\end{figure}
\begin{figure}[h]
    \centering
    \includegraphics[width=6.6cm]{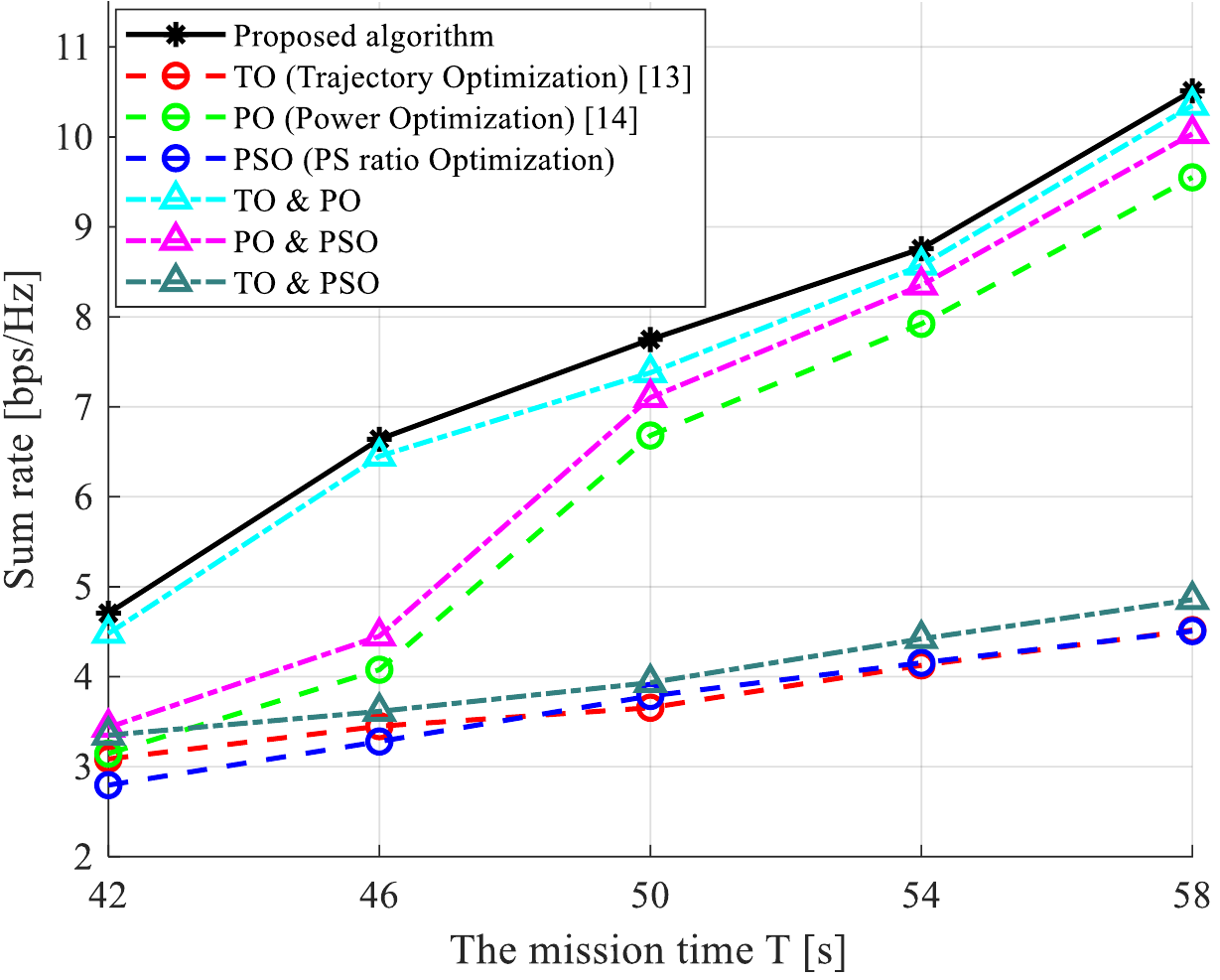}
    \caption{Sum rate versus the different mission time}
\end{figure}
\vspace{-0.2cm}

\section{Conclusions}
In this correspondence, we propose the AIRS-assisted vehicular networks systems, where the rate-splitting SWIPT receivers are assumed at vehicles. We aim at the sum-rate maximization by jointly optimizing the trajectory and phase shift design of AIRS, transmit power and rate allocation for RSMA along with PS ratio. Via simulations, the performance superiority of the proposed algorithm is verified compared to the benchmark schemes. As future works, we extend the use of AIRS for the integrated space-air-ground or space-air-sea networks.

\end{document}